\documentclass{article}

\usepackage{arxiv}

\usepackage[utf8]{inputenc} % allow utf-8 input
\usepackage[T1]{fontenc}    % use 8-bit T1 fonts
\usepackage{hyperref}       % hyperlinks
\usepackage{url}            % simple URL typesetting
\usepackage{booktabs}       % professional-quality tables
\usepackage{amsfonts}       % blackboard math symbols
\usepackage{nicefrac}       % compact symbols for 1/2, etc.
\usepackage{microtype}      % microtypography
\usepackage{lipsum}		% Can be removed after putting your text content
\usepackage{graphicx}
\usepackage{natbib}
\usepackage{doi}

\usepackage[dvipsnames]{xcolor}
\usepackage{geometry}
\usepackage{setspace}
\usepackage{pdflscape}
\usepackage{longtable}
\usepackage{rotating}
\graphicspath{ {./images/} {./figures/} }
\usepackage{tikz}
\usepackage{pgfplots}

\title{Automatic Pharma News Categorization}

\author{Stanislaw Adaszewski \\
	Pharma Research and Early Development Operations\\
	Roche Innovation Center Basel\\
	F. Hoffmann - La Roche AG\\
	Basel, CH-4070, Switzerland\\
	\texttt{stanislaw.adaszewski@roche.com} \\
	\And
	Pascal Kuner \\
	Pharma Research and Early Development Operations\\
	Roche Innovation Center Basel\\
	F. Hoffmann - La Roche AG\\
	Basel, CH-4070, Switzerland\\
	\And
	Ralf J. Jaeger \\
	Pharma Research and Early Development Operations\\
	Roche Innovation Center Basel\\
	F. Hoffmann - La Roche AG\\
	Basel, CH-4070, Switzerland\\
}

\renewcommand{\shorttitle}{Automatic Pharma News Categorization}

\hypersetup{
pdftitle={Automatic Pharma News Categorization},
pdfsubject={cs.AI},
pdfauthor={Stanislaw Adaszewski, Pascal Kuner, Ralf J. Jaeger},
pdfkeywords={natural, language processing, transformers, pharma, news, ensemble models},
}

\begin{document}
\maketitle

\begin{abstract}
	We use a text dataset consisting of 23 news categories relevant to pharma information science, in order to compare the fine-tuning performance of multiple transformer models in a classification task. Using a well-balanced dataset with multiple autoregressive and autocoding transformation models, we compare their fine-tuning performance. To validate the winning approach, we perform diagnostics of model behavior on mispredicted instances, including inspection of category-wise metrics, evaluation of prediction certainty and assessment of latent space representations. Lastly, we propose an ensemble model consisting of the top performing individual predictors and demonstrate that this approach offers a modest improvement in the F1 metric.
\end{abstract}

% keywords can be removed

\section{Introduction}
	
    Aggregation and analysis of news related to business fields of a pharma company is a cumbersome process. Users read and process individual websites, partially aggregated news feeds and items collected from various sources, as well as edited digests, commercial feeds and newsletters received by email. There are also no tools for creating and searching categorized content or for querying and retrieving content from sources defined ad hoc. Typically, users have access to different pieces of information published in various formats on distinct media platforms. Preparation of digests for sharing with other users requires manual extraction, adaptation, consolidation and grouping of articles. This is a time-consuming approach with results varying on a case-by-case basis, which leads to the demand for an improved process.
	
	To reduce quality variations and manual overhead, an automated system has been put in place to aggregate the wide spectrum of sources. The platform ensures that data are harmonized and processed into a standardized format, indexed with both internal and external terminologies and presented in a unified feed. In order to take into account individual user interests when sorting items in the presentation layer, we employ a deep learning solution.

    We considered regular expressions (RE) as an alternative. Furthermore, we used RE to bootstrap the training dataset. However, articles found by manually constructing the expressions suffered heavily from false instances (positives and negatives). For example, when searching for "clinical phases" one may use a RE like  \texttt{(C$|$clinical$\backslash$s)? ((phases?$|$stages?)$\backslash$s)([1-4IV/]\{1,3\})} to identify the relevant articles. During curation, we found false positives, in particular for stage (\texttt{[1-4IV]\{1-3\}}) related strings used with disease stages (i.e. in the wrong context) and false negatives for less commonly used expressions, like early/late, first/second phase/stage. Boolean and RE queries are prone to these types of mismatches and omissions, whereas DL approaches can build awareness of certain concepts and account for ambiguities.

	Based on a questionnaire, users inform the system of the level of interest they have in the given topics. Each of the topics is predefined by a set of curated articles. Consequently, a natural language processing (NLP) model is trained to differentiate between categories of news selected and rated by the user. This allows to pivot, filter and sort articles based on the interest scores. A presentation enriched in relevant items minimizes the time to find the required information, thereby meeting the user's needs faster and more accurately.
	
	This article describes our approach to developing an optimal production model for automated pharma news category assignment.

\section{Dataset preparation}
	
	The dataset consists of news articles grouped in 23 categories. The definitions of most were derived from interviews with stakeholders belonging to the involved working groups. Definitions consist of inclusion and exclusion criteria (e.g. exclusion of veterinary topics to keep the focus on human medicines).
	
	The articles (i.e. title plus body) of the training dataset were always assigned to a single category using the main aspect of the article. We collected only the directly accessible parts of articles, i.e. not following full text links. Therefore, articles consisting of only the title or the title and truncated content were included. The dataset shows a bi-modal distribution, where the lengths of the articles vary between 22 and 32852 characters (4 and 5338 words), with an average of 1912.40 (289.76 words), a median of 380.00 characters (59.00 words). We identified peaks around 280 and 3150 characters and 40 and 380 words, respectively. Very long articles (i.e. more than max/2 characters) are rare (5) in the training set. 
     
    The identification of matching articles and the categorization process were initiated using a set of regular expressions matching topic-specific keywords and synonyms. 
	
	For the sake of brevity, the full description of categories is provided in table \ref{tab:catdesc} in the Appendix. A single category may cover more than one topic (e.g. JobBizLaw). These categories may be split in the future to achieve more granularity if required by the users.
	
	On a high level, 21 out of the 23 categories can be grouped into 3 major areas - "Clinical Trials" (Appr/Withd/Sub, C1, C1/C2, C2, C2/C3, C3, C4 \& L-M), "Research" (Chem, Conf, DIA, DigiBioMLDev, MoA, PHC, PhDev, RWD) and "Business/Regulatory" (JobBizLaw, M\&AA, OTC, Patents, Pricing/Costs, PV-Reg). The remaining two categories are broader. The category "Other" contains all relevant Pharma topics (mainly biology topics) not included in any other classification, whereas the "OffTopic" category contains a range of examples of non-Pharma-related news which are included in relevant news streams but covering broader or more generic scientific interests (e.g. astronomy).
	
	The "Clinical Trials" categories refer to the events of beginning and end of different clinical drug development phases, as well as submissions to, approvals and withdrawals of approval by the health authorities. The "Research" categories contain news items treating recent developments in the respective research areas. Finally, the "Business/Regulatory" categories track articles related to business and law with different aspects of emphasis.
	
	Detailed statistics of the dataset are presented in table \ref{tab:dsstats} in the Appendix, including the number of examples in each category, the distribution of stopwords and non-stopwords in each category and totals for the entire dataset.
	
	Our approach to determine the number of articles per category was guided by a) defining the minimum number of articles per category as approximately 200, b) the number available in a pool consisting of recent articles, i.e. not older than one year and c) strictly avoiding strong imbalance between the categories, i.e. striving for a ratio lower than 1:4~\citep{krawczyk2016learning}. We achieved a balance ratio of 1:3.65 or better for the listed categories. In each round of curation, we split the dataset between two curators, followed by two rounds of cross-validation on large samples.

\section{Architectures}
	
	The majority of current state of the art NLP Deep Learning architectures can be traced back to the original Transformer~\citep{vaswani2017attention} model. The Transformer follows an encoder-decoder architecture. The encoder maps an input sequence of binary symbol representations to a sequence of continuous representations. The decoder then generates an output sequence of symbols one element at a time. The encoder is composed of a stack of N identical layers, each composed of a multi-head self-attention mechanism, and a position-wise fully connected network. A residual connection passes around each of the above components, followed by layer normalization. The decoder is similarly composed of N identical layers. Its layers feature a third component, which performs multi-head attention over the output of the encoder. Like in the encoder, residual connections are placed around each of the components, followed by layer normalization.
	
	NLP transformer models are frequently used in a two-step fashion. We distinguish between model pre-training and fine-tuning. Pre-training is typically performed on a very large corpus and uses the model's native task (e.g. missing token reconstruction) and loss function. Fine-tuning uses a downstream, usually smaller dataset to build a simple estimator (e.g. a classifier) while leveraging the knowledge accumulated in the transformer during the training on its native task. To this end, the architecture is modified by providing a classifier head and loss function. In this work, we consider only the fine-tuning of pre-trained NLP models in order to perform the Pharma news classification task.
	
	We use the NLP architecture implementations from the Transformers~\citep{wolf2020transformers} package. The software manages as well the retrieval of pre-trained model weights from the free and public repository at \url{https://huggingface.co/models}.
	
	We evaluate fine-tuning of multiple pre-trained models from the following families: BERT~\citep{devlin2018bert}, BART~\citep{lewis2019bart}, GPT~\citep{radford2019language}, XLnet~\citep{yang2019xlnet}, XLM~\citep{lample2019cross}, RoBERTa~\citep{liu2019roberta}, DistilBERT~\citep{sanh2019distilbert}, AlBERT~\citep{lan2019albert},  XLM-RoBERTa~\citep{conneau2019unsupervised}, Flaubert~\citep{le2019flaubert},  Electra~\citep{clark2020electra} and  Longformer~\citep{beltagy2020longformer}.

\section{Methods}

	For each pretrained model we use the corresponding tokenizer to generate embeddings for the summaries of the articles (i.e. title+body). Due to the Graphical Processing Units (GPU) memory constraints and to maintain a reasonable batch size we truncate the length of the embeddings to 128 elements, as opposed to the default, usually 512~\citep{korotkova2020exploration}. We pad the embeddings shorter than 128 elements with model-specific padding tokens up to the desired length. This setup allows us to use 32 examples per batch per GPU. We run all the procedures on 2 GPUs by means of the pytorch-lightning~\citep{falcon2019pytorch} package.
	We use train-validation-test splits of the dataset with ratios .5~:~.25~:~.25 respectively and a fixed random seed, so that the splits are identical across different pretrained models.
	The pretrained model weights are loaded from the \url{https://huggingface.co/models} repository. The pretrained classifier head is subsequently replaced with an untrained one, consisting of 23 outputs corresponding to the number of labels in our dataset.
	We use the AdamW~\citep{loshchilov2017decoupled} optimizer for all models, with a fixed learning rate of 2e-5.
	The performance of the models is evaluated on the test partition of the dataset using the accuracy metric, as well as F1, precision and recall with micro-averaging. In addition, we log the confusion matrices for debugging purposes.
	The gradient for optimization is provided by model-specific loss functions. We train for a maximum of 50 epochs with an early stopping condition of 5 epochs of no improvement in the validation F1 score.
	Model checkpoints are saved at every epoch. We use the model weights that resulted in the best validation F1 score in order to perform the measures on the test split.
	In order to assess the prediction certainty of the model on correctly and incorrectly classified examples, we perform a Monte Carlo dropout~\citep{gal2016dropout} procedure on the top-performing model.
	To visually assess the level of separation between categories in the latent space, we compute and visualize a t-SNE~\citep{hinton2002stochastic} embedding of the transformed [CLS] token representations.
	We use the top 6 best-performing individual predictor models in order to build an ensemble classifier~\citep{opitz1999popular,polikar2006ensemble,rokach2010ensemble}. The ensemble algorithm computes the histogram of individual predictions and returns the most frequently predicted label. In case of a draw, a random label among the most-voted ones is returned. We repeat the training of all the 6 top-performing individual models 10 times (with different initial weights but the same train/validation/testing splits) and compute individual F1 scores, as well as ensemble F1 scores on the test dataset.

\section{Results}

	All individual predictor models finished training early by achieving the state of no further improvement in the validation F1 score over 5 consecutive epochs. The average duration of the training was 16.34 epochs with a standard deviation of 8.86 epochs.
	We present the accuracy, F1, precision and recall metrics of the top 10 models on the test partition of the dataset in table \ref{tab:metrshort}. Full summary can be found in table \ref{tab:metrics} in the Appendix. In order to pick the top performing models we arbitrarily focus on the F1 metric as a good summary of precision and recall. The top 6 models are in order of decreasing test F1 - \textit{roberta-large}, \textit{roberta-base}, \textit{distilbert-base-uncased}, \textit{facebook/mbart-large-en-ro}, \textit{facebook/bart-large} and \textit{xlnet-large-cased}.
	We show individual F1 metrics on the test partition for all models and all categories in table \ref{tab:percat} in the Appendix.
	
\begin{figure}
		\centering
		\begin{tikzpicture}
			\node[inner sep=0](conf) at (6, 0) { \includegraphics[width=.25\textwidth]{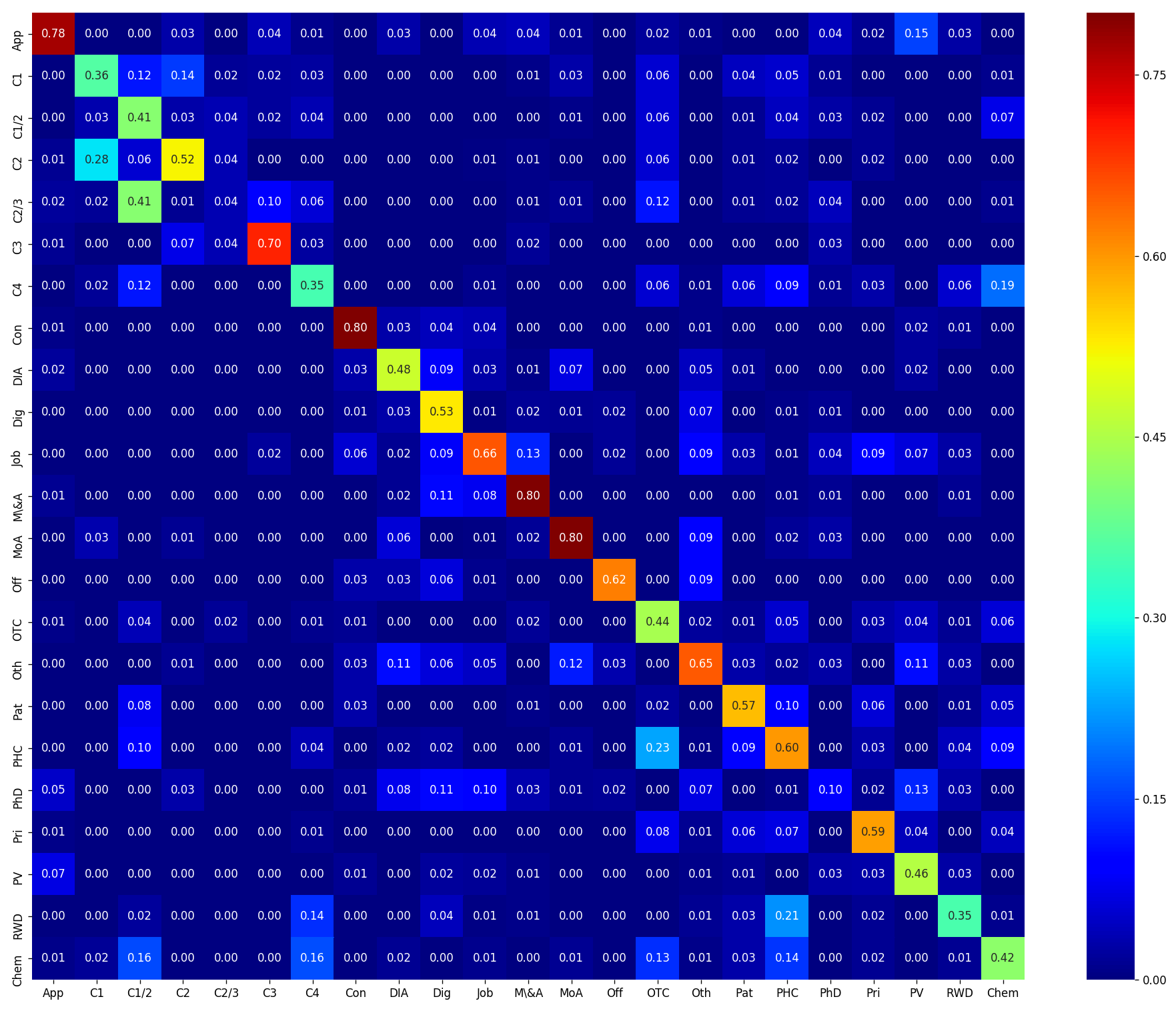} };
			\node[inner sep=0](swarm) at (6, 3.25) { \includegraphics[width=.25\textwidth]{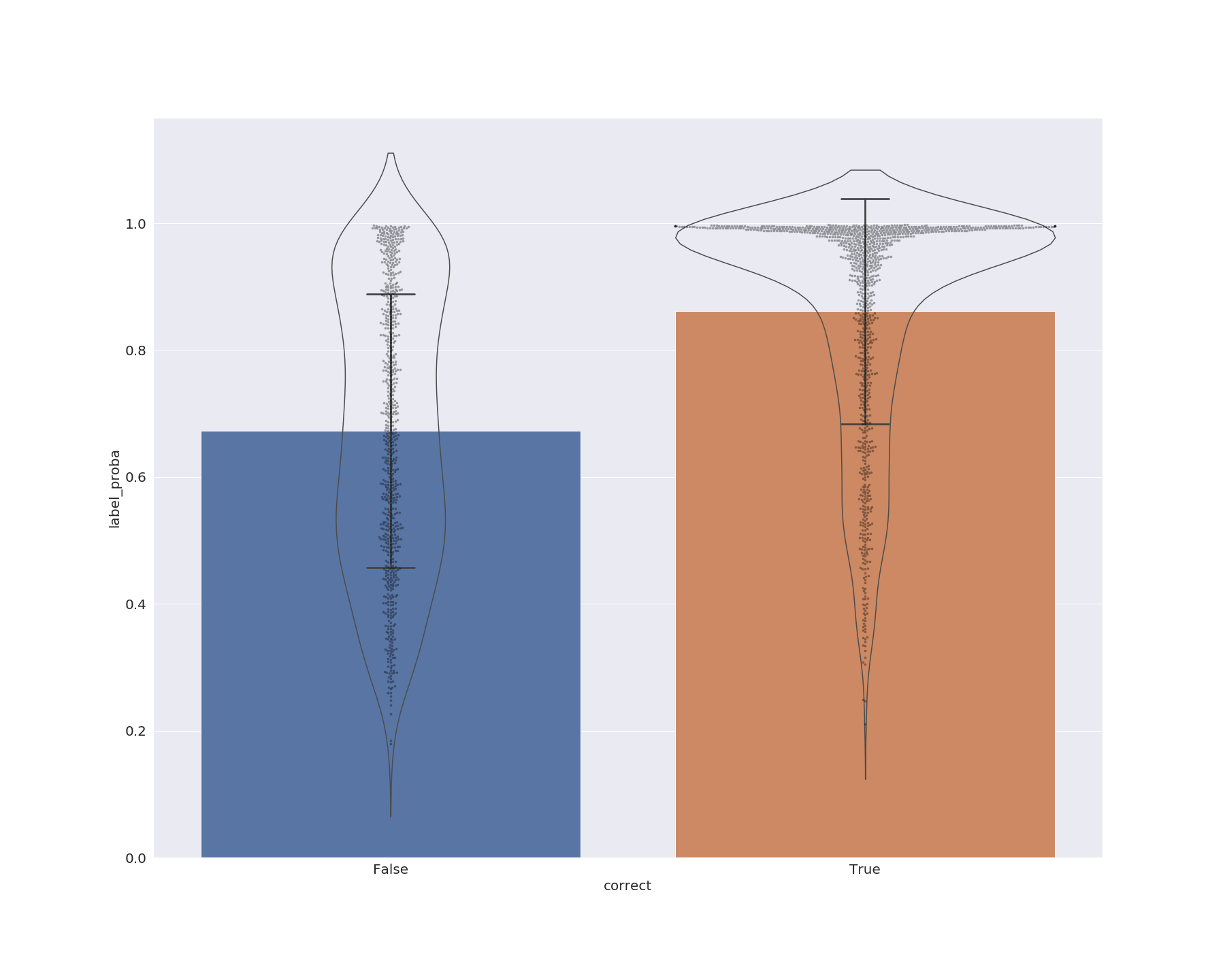} };
			\node[inner sep=0](tsne) at (0, .1\textwidth) { \includegraphics[width=.5\textwidth,height=.4\textwidth]{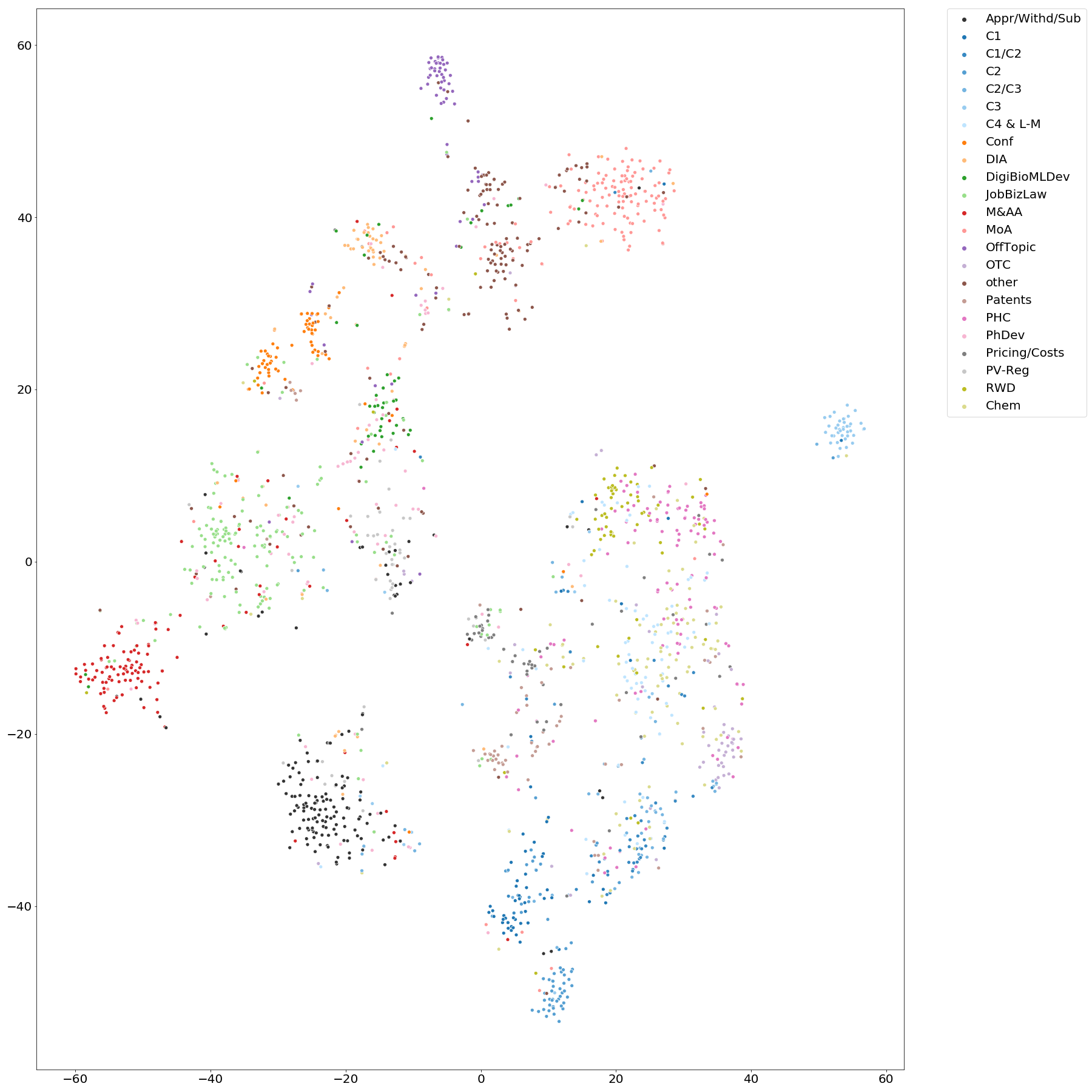} };
			\node[inner sep=0] at (-4cm,4.3cm) { A };
			\node[inner sep=0] at (8.2cm,4.3cm) { B };
			\node[inner sep=0] at (8.2cm,1.4cm) { C };
		\end{tikzpicture}
	
		\caption{A. T-SNE embedding of the last hidden state of [CLS] token in the Roberta network; B. Prediction probability distribution in incorrectly vs. correctly predicted instances; C. Normalized confusion matrix of the best model - \textit{roberta-large} - on the test split of the dataset. To obtain the normalized version, confusion matrix row values are divided by the number of examples in the respective categories.}
		\label{fig:results}
	\end{figure}
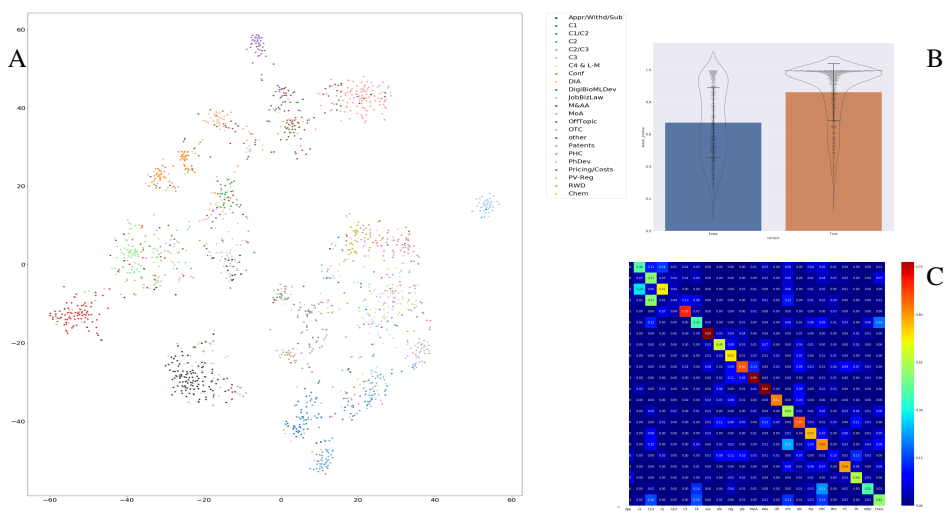

	The Monte Carlo dropout~\citep{gal2016dropout} results on the best-performing model are presented in figure \ref{fig:results} B. The prediction probabilities for incorrectly predicted instances are distributed almost uniformly between 0 an 1, whereas the probabilities for correctly predicted instances are close to 1.0 in the majority of cases. This result is expected and desired and would likely allow to achieve an even higher enrichment of articles of interest when using a confidence prediction threshold.
	
	The visualization of t-SNE~\citep{hinton2002stochastic} embeddings of the transformed [CLS] token representations from the fine-tuned \textit{roberta-large} model are presented in figure \ref{fig:results} A. Visual inspection confirms the formation of clearly separated clusters for each category. The clusters contain some problematic instances that are labeled differently than the cluster majority. This is in accord with the observed confusion matrix (figure \ref{fig:results} C), e.g. the misclassifications between M\&AA and JobBizLaw or C1 and C2.
	
	Overall, the top individual models have good performance and behave correctly in terms of prediction probability and the visual inspection of their embeddings.
	The results of the ensemble model created from the top 6 best-performing individual predictors are presented in table \ref{tab:ensemble}. As expected, the average F1 score is slightly higher than any of the individual models.
	
	\begin{table}
		\centering
		\label{tab:metrshort}
		\caption{Summary metrics for the top-10 models.}
		\begin{tabular}{ l c c c c c }	
			\toprule
			\textbf{Pretrained Model} & \textbf{Accuracy} & \textbf{F1} & \textbf{Loss} & \textbf{Precision} & \textbf{Recall} \\
			\midrule
			roberta-large &          0.57 &    0.56 &       1.4 &           0.56 &        0.57 \\
			roberta-base &          0.56 &    0.55 &       1.5 &           0.55 &        0.56 \\
			distilbert-base-uncased &          0.55 &    0.54 &       1.5 &           0.55 &        0.55 \\
			facebook/mbart-large-en-ro &          0.54 &    0.54 &       1.7 &           0.55 &        0.54 \\
			facebook/bart-large &          0.55 &    0.54 &       1.5 &           0.55 &        0.55 \\
			xlnet-large-cased &          0.55 &    0.54 &       1.7 &           0.56 &        0.55 \\
			bert-large-uncased &          0.54 &    0.53 &       1.5 &           0.54 &        0.54 \\
			facebook/bart-base &          0.54 &    0.53 &       1.6 &           0.53 &        0.54 \\
			facebook/bart-large-xsum &          0.54 &    0.53 &       1.5 &           0.54 &        0.54 \\
			xlm-mlm-en-2048 &          0.54 &    0.53 &       1.7 &           0.55 &        0.54 \\
			\bottomrule
		\end{tabular}
	\end{table}
	
	\begin{table}
		\centering
		\caption{Results of the Ensemble model testing. Test F1 scores of the individual predictors and the Ensemble model are presented, averaged over 10 runs with different initial weights but the same train/validation/test splits. M1, roberta-large; M2, roberta-base; M3, distilbert-base-uncased; M4, facebook/mbart-large-en-ro; M5, facebook/bart-large; M6, xlnet-large-cased.}
		\label{tab:ensemble}
		\begin{tabular}{lrrrrrrr}
			\toprule
			{} & \textbf{M1} & \textbf{M2} & \textbf{M3} & \textbf{M4} & \textbf{M5} & \textbf{M6} & \textbf{Ensemble} \\
			\midrule
			\textbf{F1} & 0.56 & 0.55 & 0.54 & 0.54 & 0.54 & 0.54 & 0.58 \\
			\bottomrule
		\end{tabular}
	\end{table}

\section{Discussion}

	Development of well-balanced and curated training sets is a key prerequisite for performing categorization in a consistent and correct manner. In this study, we did not investigate the effects of the bimodal distribution of number of characters/words on the inference performance. One potential venue for future research is to examine if the DL algorithm obtains better results when operating on subsets of articles grouped by length.
	
	In our current setup, the top individual models have good performance and behave correctly in the terms of prediction probability and the visual inspection of their embeddings. The ensemble model performs better than any of the individual models and has the test F1 score higher by 0.02 than the best performing individual model. We are satisfied with this level of performance and successfully deploy this setup in a production environment. Nevertheless, many interesting opportunities for improvement exist.
	
	We must consider that despite our best efforts some labels in the training dataset might have been attributed incorrectly by us. On inspection of a sample, we found a few articles which were assigned to different categories by the respective curators because of ambiguities in the texts or insufficient precision of curation rules. Another source for labeling flaws derives from articles being inherently noisy due to the overlap between certain pairs of categories. Therefore, exploring techniques such as bootstrapping~\citep{reed2014training} or loss correction~\citep{arazo2019unsupervised} might be one of the future venues to improve our model metrics. The observation of a bimodal distribution in the collected articles needs further analysis, as the semantic difference between the two groups seem to be that one contains single- and the other one - multi-topic articles. A multi-label approach at the level of articles or paragraphs/sentences could be another alternative to address the ambiguity. In particular, longer articles (i.e. more than 300-400 words, according to our observations) are at risk for mislabelling because they usually summarize several aspects of a story (e.g. event reports or articles on mergers and acquisitions) and provide a historical background or discuss future plans and prospects.
	
	Curation of the dataset by two curators showed both its advantages and limits. A slight improvement could be obtained by comparing the results from corpora created independently and/or using a crowd-sourced approach, with additional cross-checked labeling, thereby further reducing potential bias caused by the curators and optimizing the future curation process. However, this approach would be much more time- and resource consuming.
	
	Furthermore, the slight improvement offered by the ensemble model is not necessarily justifying the need to maintain 6 transformer models instead of just one. It seems reasonable that a single architecture might exist that would perform on the level of the Ensemble classifier. To this end, employing the neural architecture search approach described in~\citep{zhu2020autotrans} or more generally~\citep{pham2018efficient} could be a very interesting alternative to the ensemble approach.
	
	Lastly, we are interested in reformulating the problem in a few-shot learning setup, where the user provides a few examples of what is within the scope of his interest instead of formally declaring a category. Some of the most successful approaches in this area make use of the siamese networks~\citep{yan2018few,chicco2021siamese,reimers2019sentence,sentencetransformers}. In our scenario, we can train a model to determine for any pair of abstracts whether they belong to the same category or not. Subsequently, new articles can be evaluated in this manner against the examples provided by the user, therefore determining if they belong to a category of interest of the user without formalizing what the category is.

\section{Summary}

	We prepared a curated dataset of news articles grouped in 23 categories relevant to pharma. Subsequently, we used it to evaluate a large cross-section of fine-tuned state-of-the-art NLP transformer models in the context of a classification task. We identified the best performing architectures and pretrained weight sets and validated their behavior in terms of prediction certainty and the visual quality of their embeddings. We constructed an ensemble model consisting of the top 6 of such individual predictors and demonstrated that it outperforms the best individual model by 0.02 F1 score on the test partition of the dataset. We identified three major areas for future research - addressing the label ambiguity problem, neural architecture search and few-shot learning. The first two aim respectively at improving the model metrics and reducing the ensemble model back to a single architecture. The third reformulates the problem as an arbitrary category differentiation task eliminating altogether the need for formal category definitions.

\bibliographystyle{unsrtnat}
\bibliography{news-cat-nlp}

\newpage
	
	\appendix
	\section{Dataset Details}
	
	\begin{table}[h]
		\centering
		\caption{Dataset statistics per category and totals. Cat, category; Std, standard deviation; Min, minimum; Max, maximum; Med, median.}
		\label{tab:dsstats}
		
		\begin{tabular}{lrrrrrrrrrrr}
			\toprule
			{} & {} & \multicolumn{5}{c}{\textbf{Non-stopwords}} & \multicolumn{5}{c}{\textbf{Stopwords}} \\
			\textbf{Cat} & \textbf{Items} & \textbf{Mean} & \textbf{Std} & \textbf{Min} & \textbf{Max} & \textbf{Med} & \textbf{Mean} & \textbf{Std} & \textbf{Min} & \textbf{Max} & \textbf{Med} \\
			\midrule
			Appr/... &                 623 &              43 &             34 &              7 &            545 &                45 &          13 &         14 &          0 &        243 &            11 \\
			C1 &                 242 &             387 &            505 &             12 &           3511 &               254 &         181 &        274 &          2 &       2427 &           103 \\
			C1/C2 &                 202 &             599 &            505 &             20 &           4469 &               477 &         223 &        181 &          4 &       1447 &           184 \\
			C2 &                 274 &             310 &            469 &              9 &           4071 &                95 &         127 &        230 &          1 &       2145 &            23 \\
			C2/C3 &                 217 &             539 &            527 &             16 &           2673 &               395 &         204 &        209 &          2 &       1022 &           128 \\
			C3 &                 201 &              66 &             90 &              8 &            591 &                52 &          21 &         38 &          0 &        254 &            13 \\
			C4\&L-M &                 319 &             669 &            663 &             26 &           4720 &               471 &         259 &        273 &          9 &       1680 &           174 \\
			Chem &                 385 &             457 &            390 &             24 &           4469 &               337 &         170 &        150 &          7 &       1447 &           128 \\
			Conf &                 279 &              35 &             52 &              6 &            598 &                20 &          10 &         27 &          0 &        388 &             5 \\
			DIA &                 261 &              32 &             50 &              6 &            778 &                27 &          10 &         18 &          0 &        270 &             8 \\
			DigiBio... &                 189 &              48 &             87 &              8 &           1102 &                40 &          18 &         46 &          1 &        610 &            13 \\
			JobBiz... &                 665 &              44 &             66 &              5 &           1511 &                38 &          16 &         36 &          0 &        838 &            13 \\
			M\&AA &                 477 &              42 &             34 &              8 &            605 &                40 &          14 &         15 &          0 &        249 &            13 \\
			MoA &                 564 &              44 &             83 &              5 &            726 &                27 &          17 &         36 &          0 &        309 &            10 \\
			OTC &                 208 &             761 &            743 &             27 &           4597 &               506 &         276 &        276 &          0 &       1552 &           179 \\
			PHC &                 448 &             610 &            561 &             21 &           4166 &               453 &         238 &        230 &          0 &       1420 &           161 \\
			PV-Reg &                 182 &              46 &             72 &              5 &            774 &                39 &          18 &         42 &          0 &        496 &            12 \\
			Patents &                 267 &             576 &            631 &              8 &           4166 &               434 &         202 &        235 &          0 &       1816 &           143 \\
			PhDev &                 288 &              35 &             46 &              5 &            753 &                33 &          13 &         29 &          0 &        490 &            11 \\
			Pricing/... &                 254 &             432 &            351 &              7 &           1895 &               451 &         143 &        125 &          0 &        784 &           130 \\
			RWD &                 285 &             618 &            520 &             10 &           4061 &               466 &         260 &        251 &          1 &       1623 &           183 \\
			OffTopic &                 244 &              42 &             57 &              5 &            519 &                34 &          18 &         31 &          0 &        287 &            13 \\
			Other &                 612 &              43 &             65 &              5 &            696 &                32 &          19 &         35 &          0 &        417 &            14 \\
			\textbf{Total} &                \textbf{7686} &             \textbf{244} &            \textbf{431} &              \textbf{5} &           \textbf{4720} &                \textbf{51} &          \textbf{93} &        \textbf{176} &          \textbf{0} &       \textbf{2427} &            \textbf{18} \\
			\bottomrule
		\end{tabular}
	\end{table}

	\newgeometry{top=1cm, bottom=1cm, left=1cm, right=1cm}
	\begin{landscape}
		\pagestyle{empty}
		\renewcommand{\arraystretch}{2.5}
		\setlength{\baselineskip}{0pt}
		\centering
		\begin{spacing}{0.4}
			\begin{longtable}{ c p{8cm} p{15cm} }
				\caption{Description of the categories in our dataset.}
				\label{tab:catdesc} \\
				\toprule
				\textbf{\#} & \textbf{Group /} \textbf{Name} \textbf{(Abbreviation)} & \textbf{Description} \\
				\midrule
				
				\textbf{1} & Clinical Trials / Approvals, Withdrawals, Submissions (Appr/Withd/Sub) & Covers all kinds of regulatory relevant actions beginning from registration of a drug name, entry in to human, through development and marketing up to retraction of a drug. \\
				
				\textbf{2} & Clinical Trials / Clinical Phase 1 (C1) & Refers to Clinical phase 1 and subphases thereof \\
				
				\textbf{3} & Clinical Trials / Clinical Phase 1/2 (C1/C2) & Refers to Clinical phase 1\&2 combined trials. \\
				
				\textbf{4} & Clinical Trials / Clinical Phase 2 (C2) & Refers to Clinical phase 2 and subphases thereof. \\
				
				\textbf{5} & Clinical Trials / Clinical Phase 2/3 (C2/C3) & Refers to Clinical phase 2\&3 combined trials. \\
				
				\textbf{6} & Clinical Trials / Clinical Phase 3 (C3) & Refers to Clinical phase 3 and subphases thereof. \\
				
				\textbf{7} & Clinical Trials / Clinical Phase 4 \& L-M (C4 \& L-M) & Refers to launch, marketing, postmarketing, and C4 postmarkting trial topics \\
				
				\textbf{8} & Research / Conferences (Conf) & Includes information on upcoming, about or of content presented at conferences. \\
				
				\textbf{9} & Research / Diagnostics (DIA) & Very generically covers diagnotics related information. \\
				
				\textbf{10} & Research / Digital Biomarkers and Machine Learning (DigiBioMLDev) & Includes news about new digital biomarker, the development thereof and methods or tools used in this context. \\
				
				\textbf{11} & Business / Jobs, Business, Law (JobBizLaw) & Contains news related to pharma business job roles, staff or budget related news, company development and non patent lawsuits. \\
				
				\textbf{12} & Business / Mergers \& Acquisitions and Agreements (M\&AA) & Contains news on merger \& acquisitions and agreements in the pharma and disagnostic field. \\
				
				\textbf{13} & Research / Mode / Mechanism of Action (MoA) & Covers news on mechanism/mode of action and related preclinical topics e.g. toxicity and drug safety topics. \\
				
				\textbf{14} & - / OffTopic, i.e. not pharma, biology, diagnostics related (OffTopic) & Holds news unrelated to Pharma/Dia topics (e.g. astronomy)  but included for some reason in the news streams. \\
				
				\textbf{15} & Business / Over-the-counter (OTC) & Pertains to news on over the counter drugs and generics. \\
				
				\textbf{16} & - / Pharma topics not covered by any other category (Other) & Includes news related to Pharma/DIA or biology (evolution, veterinary, etc). \\
				
				\textbf{17} & Business / Patent applications and patent related lawsuits (Patents) & Contains news on patents, IP processes, claim and patent lawsuits. \\
				
				\textbf{18} & Research / Personalized healthcare, precision medicine (PHC) & Includes news on patient stratification and personalized health care. \\
				
				\textbf{19} & Research / Pharma Development (PhDev) & Contains Pharma development news like trial design, strategies, (not clinicial C1-C4) administrative processes and tools. \\
				
				\textbf{20} & Business / Drug Pricing, Costs, Reimbursements (Pricing/Costs) & Categorizes news on topics related to drug pricing, costs, insurance and reimbursement. \\ 
				
				\textbf{21} & Business / Pharmacovigilance, Regulatory (PV-Reg) & Includes news on pharmacovigilance, regulatory warnings like "dear doctor" letters and other regulatory information. \\
				
				\textbf{22} & Research / Real world data and Real world evidence (RWD) & Holds news on real world data and clinical or social media data repositories. \\
				
				\textbf{23} & Research / Chemistry (Chem) & Includes news on chemical compounds, reactions, methods, data etc. \\
				
				\bottomrule
			\end{longtable}
		\end{spacing}
	\end{landscape}
	\restoregeometry
	
	\section{Results Details}
	
	\renewcommand{\arraystretch}{0.5}
	\newgeometry{top=1cm, bottom=1cm, left=1cm, right=1cm}
	\begin{landscape}
		\centering
		\pagestyle{empty}
		\begin{longtable}{ l c c c c c }	
			\caption{Table with metrics for each model}
			\label{tab:metrics}	\\
			\toprule
			\textbf{pretrained\_model} & \textbf{test\_accuracy} & \textbf{test\_f1} & \textbf{test\_loss} & \textbf{test\_precision} & \textbf{test\_recall} \\
			\midrule
			LitNLPClassifier &             - &       - &         - &              - &           - \\
			albert-base-v1 &          0.52 &    0.51 &       1.6 &           0.52 &        0.52 \\
			albert-base-v2 &          0.36 &    0.35 &       2.1 &           0.37 &        0.36 \\
			albert-large-v1 &          0.48 &    0.48 &       1.7 &           0.48 &        0.48 \\
			albert-large-v2 &          0.26 &    0.21 &       2.3 &           0.23 &        0.26 \\
			albert-xlarge-v1 &          0.12 &   0.043 &       2.8 &          0.034 &        0.12 \\
			albert-xlarge-v2 &          0.46 &    0.45 &       1.7 &           0.47 &        0.46 \\
			albert-xxlarge-v1 &          0.35 &    0.32 &       2.3 &           0.33 &        0.35 \\
			albert-xxlarge-v2 &          0.31 &    0.28 &       2.2 &           0.29 &        0.31 \\
			allenai/longformer-base-4096 &             - &       - &         - &              - &           - \\
			allenai/longformer-base-4096-extra.pos.embd.only &             - &       - &         - &              - &           - \\
			allenai/longformer-large-4096 &             - &       - &         - &              - &           - \\
			allenai/longformer-large-4096-extra.pos.embd.only &             - &       - &         - &              - &           - \\
			allenai/longformer-large-4096-finetuned-triviaqa &             - &       - &         - &              - &           - \\
			bert-base-cased &          0.52 &    0.52 &       1.5 &           0.52 &        0.52 \\
			bert-base-multilingual-cased &          0.47 &    0.45 &       1.7 &           0.46 &        0.47 \\
			bert-base-multilingual-uncased &          0.47 &    0.45 &       1.7 &           0.45 &        0.47 \\
			bert-base-uncased &          0.53 &    0.52 &       1.5 &           0.52 &        0.53 \\
			bert-large-cased &          0.53 &    0.52 &       1.5 &           0.52 &        0.53 \\
			bert-large-cased-whole-word-masking &          0.53 &    0.51 &       1.5 &           0.53 &        0.53 \\
			bert-large-cased-whole-word-masking-finetuned-... &          0.53 &    0.52 &       1.6 &           0.51 &        0.53 \\
			bert-large-uncased &          0.54 &    0.53 &       1.5 &           0.54 &        0.54 \\
			bert-large-uncased-whole-word-masking &          0.54 &    0.52 &       1.5 &           0.54 &        0.54 \\
			bert-large-uncased-whole-word-masking-finetune... &          0.48 &    0.47 &       1.7 &           0.49 &        0.48 \\
			camembert-base &          0.45 &    0.42 &       1.9 &           0.43 &        0.45 \\
			distilbert-base-cased &          0.49 &    0.49 &       1.7 &           0.51 &        0.49 \\
			distilbert-base-cased-distilled-squad &          0.46 &    0.46 &       1.7 &           0.47 &        0.46 \\
			distilbert-base-multilingual-cased &          0.48 &    0.46 &       1.7 &           0.47 &        0.48 \\
			distilbert-base-uncased &          0.55 &    0.54 &       1.5 &           0.55 &        0.55 \\
			distilbert-base-uncased-distilled-squad &          0.49 &    0.48 &       1.6 &            0.5 &        0.49 \\
			distilroberta-base &          0.51 &     0.5 &       1.5 &           0.51 &        0.51 \\
			facebook/bart-base &          0.54 &    0.53 &       1.6 &           0.53 &        0.54 \\
			facebook/bart-large &          0.55 &    0.54 &       1.5 &           0.55 &        0.55 \\
			facebook/bart-large-cnn &           0.5 &    0.49 &       1.6 &           0.52 &         0.5 \\
			facebook/bart-large-xsum &          0.54 &    0.53 &       1.5 &           0.54 &        0.54 \\
			facebook/mbart-large-en-ro &          0.54 &    0.54 &       1.7 &           0.55 &        0.54 \\
			flaubert/flaubert\_base\_cased &           0.5 &     0.5 &       1.6 &           0.52 &         0.5 \\
			flaubert/flaubert\_base\_uncased &          0.48 &    0.48 &       1.8 &           0.49 &        0.48 \\
			flaubert/flaubert\_large\_cased &          0.48 &    0.46 &       1.8 &           0.49 &        0.48 \\
			flaubert/flaubert\_small\_cased &          0.45 &    0.44 &       1.9 &           0.45 &        0.45 \\
			google/electra-base-discriminator &          0.49 &    0.47 &       1.7 &           0.48 &        0.49 \\
			google/electra-base-generator &           0.4 &    0.36 &         2 &           0.36 &         0.4 \\
			google/electra-large-discriminator &          0.49 &    0.47 &       1.8 &           0.48 &        0.49 \\
			google/electra-large-generator &          0.45 &    0.39 &       1.9 &            0.4 &        0.45 \\
			google/electra-small-discriminator &           0.4 &    0.34 &       1.9 &           0.34 &         0.4 \\
			google/electra-small-generator &          0.45 &    0.45 &       1.9 &           0.47 &        0.45 \\
			roberta-base &          0.56 &    0.55 &       1.5 &           0.55 &        0.56 \\
			roberta-base-openai-detector &          0.48 &    0.47 &       1.8 &            0.5 &        0.48 \\
			roberta-large &          0.57 &    0.56 &       1.4 &           0.56 &        0.57 \\
			roberta-large-mnli &           0.5 &    0.49 &       1.6 &           0.51 &         0.5 \\
			roberta-large-openai-detector &          0.53 &    0.52 &       1.5 &           0.53 &        0.53 \\
			xlm-clm-ende-1024 &          0.49 &    0.47 &       1.7 &            0.5 &        0.49 \\
			xlm-clm-enfr-1024 &          0.44 &    0.43 &       1.9 &           0.46 &        0.44 \\
			xlm-mlm-100-1280 &           0.5 &    0.48 &       1.7 &           0.52 &         0.5 \\
			xlm-mlm-17-1280 &          0.48 &    0.47 &       1.7 &            0.5 &        0.48 \\
			xlm-mlm-en-2048 &          0.54 &    0.53 &       1.7 &           0.55 &        0.54 \\
			xlm-mlm-ende-1024 &          0.49 &    0.48 &       1.7 &            0.5 &        0.49 \\
			xlm-mlm-enfr-1024 &          0.48 &    0.46 &       1.9 &           0.46 &        0.48 \\
			xlm-mlm-enro-1024 &          0.48 &    0.47 &       1.8 &           0.48 &        0.48 \\
			xlm-mlm-tlm-xnli15-1024 &         0.086 &   0.014 &       3.1 &         0.0075 &       0.086 \\
			xlm-mlm-xnli15-1024 &          0.15 &   0.076 &       2.6 &          0.055 &        0.15 \\
			xlm-roberta-base &          0.51 &    0.48 &       1.5 &            0.5 &        0.51 \\
			xlm-roberta-large &          0.54 &    0.52 &       1.5 &           0.54 &        0.54 \\
			xlm-roberta-large-finetuned-conll03-english &          0.52 &    0.51 &       1.6 &           0.52 &        0.52 \\
			xlnet-base-cased &          0.53 &    0.52 &       1.5 &           0.54 &        0.53 \\
			xlnet-large-cased &          0.55 &    0.54 &       1.7 &           0.56 &        0.55 \\
			yjernite/bart\_eli5 &          0.53 &    0.52 &       1.7 &           0.53 &        0.53 \\
			\bottomrule
		\end{longtable}
	\end{landscape}
	\restoregeometry
	
	\newgeometry{top=1cm, bottom=1cm, left=1cm, right=1cm}
	\begin{landscape}
		\pagestyle{empty}
		\centering
		\footnotesize
		\setlength{\tabcolsep}{1pt}
		\setstretch{0.7}
		\renewcommand{\arraystretch}{1.0}
		\begin{longtable}{p{2.5cm} lllllllllllllllllllllll}
			\caption{Table with F1 metrics per model per category}
			\label{tab:percat}	\\
			\toprule
			\textbf{pretrained\_model} & \rotatebox{90}{\textbf{Appr/Withd/Sub}} &     \rotatebox{90}{\textbf{C1}} &  \rotatebox{90}{\textbf{C1/C2}} &    \rotatebox{90}{\textbf{C2}} &  \rotatebox{90}{\textbf{C2/C3}} &    \rotatebox{90}{\textbf{C3}} & \rotatebox{90}{\textbf{C4 \& L-M}} &   \rotatebox{90}{\textbf{Conf}} &    \rotatebox{90}{\textbf{DIA}} & \rotatebox{90}{\textbf{DigiBioMLDev}} & \rotatebox{90}{\textbf{JobBizLaw}} &  \rotatebox{90}{\textbf{M\&AA}} &   \rotatebox{90}{\textbf{MoA}} & \rotatebox{90}{\textbf{OffTopic}} &    \rotatebox{90}{\textbf{OTC}} & \rotatebox{90}{\textbf{other}} & \rotatebox{90}{\textbf{Patents}} &   \rotatebox{90}{\textbf{PHC}} &  \rotatebox{90}{\textbf{PhDev}} & \rotatebox{90}{\textbf{Pricing/Costs}} & \rotatebox{90}{\textbf{PV-Reg}} &    \rotatebox{90}{\textbf{RWD}} &   \rotatebox{90}{\textbf{Chem}} \\
			\midrule
			Lit &              - &      - &      - &     - &      - &     - &        - &      - &      - &            - &         - &     - &     - &        - &      - &     - &       - &     - &      - &             - &      - &      - &      - \\
			alb-bas-v1 &           0.78 &   0.43 &   0.39 &  0.34 &   0.35 &  0.59 &      0.3 &   0.78 &   0.36 &         0.46 &      0.53 &  0.71 &  0.73 &     0.74 &   0.21 &  0.61 &    0.43 &  0.44 &   0.11 &          0.43 &    0.4 &    0.3 &   0.39 \\
			alb-bas-v2 &           0.65 &   0.31 &   0.37 &  0.28 &   0.11 &  0.38 &     0.24 &   0.54 &  0.048 &         0.13 &      0.41 &  0.48 &  0.58 &     0.55 &   0.24 &  0.32 &     0.3 &  0.17 &  0.096 &          0.39 &  0.041 &   0.36 &   0.25 \\
			alb-lar-v1 &           0.73 &   0.27 &   0.23 &  0.34 &   0.25 &  0.59 &     0.34 &   0.72 &   0.42 &         0.41 &      0.55 &  0.65 &  0.66 &     0.71 &    0.3 &  0.56 &    0.34 &  0.35 &   0.21 &          0.43 &   0.41 &   0.36 &   0.31 \\
			alb-lar-v2 &           0.36 &   0.23 &    0.2 &  0.25 &   0.14 &   NaN &     0.23 &    NaN &    NaN &          NaN &      0.33 &  0.15 &  0.39 &      NaN &   0.16 &  0.17 &     0.2 &  0.31 &    NaN &           0.4 &    NaN &   0.27 &   0.34 \\
			alb-xla-v1 &          0.012 &    NaN &    NaN &   NaN &    NaN &   NaN &     0.16 &    NaN &    NaN &          NaN &      0.21 &   NaN &   NaN &      NaN &    NaN &   NaN &     NaN &  0.19 &    NaN &           NaN &    NaN &    NaN &   0.12 \\
			alb-xla-v2 &           0.72 &    0.2 &   0.19 &  0.44 &   0.16 &  0.59 &    0.068 &   0.72 &   0.44 &         0.41 &      0.57 &  0.65 &   0.7 &     0.72 &    NaN &  0.49 &    0.36 &  0.25 &   0.21 &          0.43 &   0.47 &   0.29 &   0.31 \\
			alb-xxl-v1 &           0.56 &    0.3 &   0.34 &  0.28 &   0.25 &  0.29 &     0.35 &   0.46 &  0.024 &          NaN &      0.37 &  0.42 &  0.48 &    0.032 &    0.3 &  0.33 &     0.3 &  0.25 &  0.039 &          0.39 &    NaN &   0.42 &   0.23 \\
			alb-xxl-v2 &           0.53 &   0.12 &   0.13 &  0.08 &    NaN &  0.22 &    0.056 &   0.55 &  0.022 &         0.17 &      0.41 &  0.29 &  0.47 &     0.52 &  0.033 &  0.36 &    0.07 &  0.33 &   0.12 &          0.32 &  0.067 &   0.19 &    0.3 \\
			bar &           0.79 &   0.36 &   0.27 &  0.49 &    NaN &  0.57 &     0.42 &   0.73 &   0.51 &          0.5 &      0.58 &  0.77 &  0.72 &     0.73 &   0.32 &  0.59 &    0.47 &  0.48 &   0.11 &          0.48 &   0.43 &   0.33 &   0.37 \\
			bar-bas &           0.76 &   0.36 &   0.37 &  0.46 &  0.088 &  0.56 &     0.37 &   0.79 &   0.47 &         0.44 &      0.61 &  0.74 &  0.74 &     0.68 &   0.32 &  0.56 &    0.53 &   0.4 &   0.14 &           0.5 &   0.36 &   0.48 &   0.45 \\
			bar-lar &           0.77 &   0.45 &   0.32 &   0.5 &   0.29 &  0.65 &     0.32 &   0.74 &   0.43 &          0.5 &       0.6 &  0.73 &  0.68 &      0.7 &   0.36 &  0.56 &    0.51 &  0.48 &   0.12 &          0.57 &   0.56 &    0.4 &   0.41 \\
			bar-lar-cnn &           0.75 &   0.25 &   0.21 &  0.26 &   0.15 &   0.5 &     0.27 &   0.73 &   0.37 &         0.53 &      0.55 &   0.7 &  0.69 &      0.7 &   0.27 &  0.59 &    0.43 &  0.44 &   0.14 &          0.53 &    0.5 &   0.35 &   0.36 \\
			bar-lar-xsu &            0.8 &   0.37 &  0.067 &  0.44 &   0.32 &  0.56 &     0.35 &   0.77 &   0.48 &          0.5 &      0.57 &  0.72 &  0.67 &     0.76 &   0.38 &  0.57 &    0.47 &  0.49 &   0.16 &           0.5 &   0.51 &    0.4 &   0.41 \\
			ber-bas-cas &           0.77 &   0.29 &   0.29 &  0.52 &   0.35 &  0.76 &     0.36 &   0.74 &   0.36 &         0.39 &      0.54 &  0.66 &  0.72 &     0.73 &   0.21 &  0.55 &    0.52 &  0.45 &   0.16 &          0.57 &   0.44 &   0.36 &   0.37 \\
			ber-bas-mul-cas &           0.74 &    0.2 &   0.29 &   0.4 &  0.068 &  0.47 &     0.13 &   0.71 &   0.29 &         0.27 &      0.54 &  0.69 &  0.67 &     0.73 &    0.2 &  0.56 &    0.41 &  0.31 &   0.11 &          0.43 &   0.33 &   0.31 &   0.36 \\
			ber-bas-mul-unc &           0.77 &   0.29 &   0.22 &  0.33 &   0.22 &   0.5 &     0.26 &   0.72 &   0.19 &         0.42 &      0.55 &  0.68 &  0.65 &     0.69 &   0.21 &  0.55 &    0.31 &  0.41 &  0.021 &          0.46 &   0.34 &    NaN &   0.36 \\
			ber-bas-unc &           0.75 &   0.38 &   0.34 &  0.56 &   0.23 &  0.71 &     0.32 &   0.75 &   0.45 &          0.4 &      0.55 &  0.66 &  0.71 &      0.7 &   0.21 &  0.56 &    0.53 &  0.44 &   0.14 &          0.49 &   0.38 &   0.41 &   0.48 \\
			ber-lar-cas &           0.78 &   0.36 &   0.37 &  0.43 &   0.18 &  0.61 &     0.29 &   0.73 &   0.51 &          0.4 &       0.6 &   0.7 &  0.73 &     0.73 &   0.26 &  0.61 &    0.48 &  0.39 &   0.13 &          0.53 &   0.46 &   0.35 &   0.42 \\
			ber-lar-cas-who-wor-mas &           0.76 &   0.38 &   0.32 &  0.49 &  0.062 &  0.61 &     0.29 &   0.77 &   0.49 &         0.46 &      0.57 &  0.74 &  0.66 &     0.71 &   0.26 &  0.58 &    0.54 &  0.45 &  0.083 &          0.46 &   0.37 &   0.44 &   0.39 \\
			ber-lar-cas-who-wor-mas-fin-squ &           0.76 &   0.29 &   0.31 &  0.46 &   0.13 &  0.55 &     0.34 &   0.79 &   0.45 &         0.45 &      0.62 &  0.74 &  0.73 &     0.68 &   0.36 &  0.54 &    0.48 &  0.44 &   0.16 &          0.52 &   0.45 &   0.27 &   0.38 \\
			ber-lar-unc &           0.75 &    0.4 &   0.37 &  0.52 &   0.28 &  0.67 &     0.32 &   0.72 &   0.45 &         0.43 &       0.6 &  0.74 &  0.74 &     0.75 &    0.4 &  0.59 &    0.44 &  0.41 &   0.12 &           0.5 &   0.51 &   0.44 &   0.38 \\
			ber-lar-unc-who-wor-mas &           0.75 &   0.24 &    0.4 &  0.45 &  0.033 &  0.66 &     0.28 &   0.78 &   0.47 &          0.5 &      0.59 &  0.72 &  0.73 &     0.68 &   0.41 &  0.58 &    0.47 &  0.42 &   0.24 &          0.52 &   0.44 &   0.41 &   0.39 \\
			ber-lar-unc-who-wor-mas-fin-squ &            0.8 &   0.35 &   0.32 &  0.38 &  0.065 &   0.6 &     0.23 &   0.75 &   0.25 &          0.3 &      0.57 &  0.72 &  0.66 &     0.61 &   0.26 &  0.52 &    0.35 &  0.38 &    0.1 &          0.44 &   0.42 &   0.35 &   0.37 \\
			cam-bas &           0.74 &   0.17 &   0.26 &  0.19 &  0.026 &  0.64 &    0.059 &   0.71 &    0.4 &         0.26 &      0.54 &  0.69 &  0.62 &     0.51 &    NaN &  0.49 &    0.26 &  0.37 &   0.18 &          0.48 &   0.28 &   0.25 &   0.38 \\
			dis-bas &           0.72 &   0.28 &   0.29 &  0.52 &   0.33 &  0.59 &     0.34 &   0.75 &   0.33 &         0.39 &      0.59 &  0.69 &  0.66 &     0.68 &   0.36 &  0.52 &    0.44 &  0.41 &   0.13 &           0.5 &   0.34 &   0.41 &   0.46 \\
			dis-bas-cas &           0.77 &   0.34 &   0.38 &  0.51 &   0.21 &  0.65 &     0.32 &   0.67 &   0.39 &         0.29 &      0.56 &  0.63 &  0.64 &     0.67 &   0.34 &  0.49 &     0.5 &  0.41 &   0.22 &          0.53 &    0.3 &   0.37 &   0.37 \\
			dis-bas-cas-dis-squ &           0.73 &   0.24 &   0.29 &  0.44 &   0.25 &  0.49 &      0.3 &   0.66 &    0.4 &         0.42 &       0.5 &  0.67 &  0.66 &     0.64 &   0.27 &  0.51 &    0.39 &  0.32 &   0.17 &          0.39 &   0.34 &   0.31 &   0.29 \\
			dis-bas-mul-cas &           0.78 &   0.32 &   0.33 &  0.48 &  0.088 &  0.55 &     0.28 &   0.75 &   0.32 &         0.35 &      0.52 &  0.66 &  0.67 &     0.74 &    NaN &  0.51 &    0.33 &  0.34 &   0.21 &          0.44 &    0.2 &   0.37 &   0.35 \\
			dis-bas-unc &           0.77 &   0.51 &   0.31 &  0.63 &   0.13 &  0.78 &     0.28 &   0.78 &   0.43 &         0.45 &      0.59 &  0.69 &  0.73 &     0.73 &   0.39 &   0.6 &    0.55 &  0.42 &   0.29 &          0.56 &   0.44 &    0.4 &   0.38 \\
			dis-bas-unc-dis-squ &           0.76 &   0.32 &   0.31 &  0.48 &   0.18 &  0.66 &     0.32 &   0.71 &   0.44 &         0.35 &      0.55 &  0.66 &  0.68 &     0.68 &   0.13 &  0.51 &    0.37 &  0.38 &   0.13 &          0.44 &   0.42 &   0.39 &   0.37 \\
			ele-bas-dis &           0.76 &   0.44 &   0.27 &   0.4 &   0.24 &  0.56 &     0.17 &   0.78 &   0.16 &          0.4 &      0.55 &  0.68 &   0.7 &     0.74 &    NaN &  0.56 &    0.31 &  0.38 &    0.2 &          0.45 &   0.25 &   0.35 &   0.43 \\
			ele-bas-gen &           0.71 &    NaN &  0.038 &  0.45 &   0.22 &  0.57 &      NaN &   0.35 &   0.24 &          NaN &       0.5 &   0.5 &  0.63 &     0.57 &    NaN &  0.45 &    0.26 &  0.43 &   0.14 &          0.13 &   0.18 &   0.23 &   0.32 \\
			ele-lar-dis &           0.75 &   0.49 &   0.34 &  0.46 &   0.38 &  0.73 &     0.42 &   0.64 &   0.13 &         0.37 &      0.49 &  0.66 &  0.65 &      NaN &   0.33 &  0.51 &    0.42 &  0.53 &  0.083 &          0.49 &    0.2 &   0.39 &   0.31 \\
			ele-lar-gen &           0.76 &    NaN &    NaN &  0.43 &    NaN &  0.73 &      0.1 &   0.59 &    NaN &          NaN &      0.54 &  0.73 &  0.67 &     0.72 &    NaN &  0.49 &     0.2 &  0.37 &    NaN &          0.47 &   0.28 &    NaN &   0.36 \\
			ele-sma-dis &           0.63 &   0.28 &   0.14 &  0.34 &    NaN &  0.53 &    0.024 &   0.52 &  0.028 &          NaN &      0.48 &  0.51 &  0.65 &      NaN &    NaN &  0.44 &    0.25 &  0.41 &    NaN &           0.5 &    NaN &   0.11 &   0.39 \\
			ele-sma-gen &           0.73 &   0.24 &   0.35 &  0.41 &   0.17 &  0.53 &     0.34 &   0.65 &   0.37 &         0.26 &      0.37 &  0.51 &  0.63 &     0.65 &   0.18 &  0.51 &    0.48 &  0.41 &   0.25 &          0.33 &   0.33 &   0.41 &   0.39 \\
			fla &           0.75 &    0.3 &   0.28 &  0.25 &    NaN &  0.57 &     0.36 &   0.61 &   0.45 &         0.47 &      0.56 &  0.65 &  0.71 &     0.76 &   0.28 &  0.57 &    0.31 &   0.3 &  0.071 &          0.43 &   0.41 &   0.29 &   0.33 \\
			fla &           0.78 &   0.33 &   0.32 &  0.48 &   0.22 &  0.58 &     0.39 &   0.71 &   0.47 &          0.5 &      0.54 &  0.65 &  0.68 &     0.71 &   0.32 &  0.53 &    0.39 &  0.35 &   0.19 &          0.49 &   0.42 &   0.32 &   0.35 \\
			fla &           0.71 &   0.36 &   0.24 &  0.58 &   0.21 &  0.75 &     0.29 &   0.62 &    0.3 &         0.45 &      0.49 &  0.54 &   0.6 &     0.61 &   0.24 &  0.46 &     0.3 &  0.32 &   0.18 &          0.41 &   0.35 &   0.35 &    0.3 \\
			fla &           0.71 &   0.24 &   0.18 &  0.47 &   0.22 &  0.64 &     0.27 &   0.74 &   0.48 &         0.45 &      0.54 &  0.63 &   0.7 &     0.69 &   0.27 &  0.53 &    0.39 &  0.34 &   0.25 &          0.49 &   0.36 &   0.26 &   0.33 \\
			lon-bas-409 &          0.066 &    NaN &    NaN &   NaN &    NaN &   NaN &      NaN &   0.19 &    NaN &          NaN &       NaN &   NaN &   NaN &      NaN &    NaN &   NaN &     NaN &   NaN &    NaN &           NaN &    NaN &    NaN &    NaN \\
			lon-bas-409-ext &            NaN &    NaN &    NaN &   NaN &    NaN &   NaN &      NaN &  0.025 &    NaN &          NaN &       NaN &   NaN &   NaN &      NaN &   0.13 &   NaN &     NaN &   NaN &    NaN &           NaN &    NaN &  0.037 &    NaN \\
			lon-lar-409 &            NaN &  0.054 &    NaN &   NaN &    NaN &   NaN &      NaN &    NaN &    NaN &        0.047 &       NaN &   NaN &   NaN &      NaN &    NaN &   NaN &     NaN &   NaN &    NaN &           NaN &    NaN &    NaN &    NaN \\
			lon-lar-409-ext &            NaN &    NaN &    NaN &   NaN &    NaN &   NaN &      NaN &    NaN &    NaN &          NaN &      0.13 &   NaN &   NaN &      NaN &    NaN &   NaN &     NaN &   NaN &    NaN &           NaN &    NaN &    NaN &    NaN \\
			lon-lar-409-fin-tri &            NaN &    NaN &    NaN &   NaN &  0.053 &   NaN &      NaN &    NaN &    NaN &          NaN &       NaN &   NaN &   NaN &      NaN &    NaN &   NaN &     NaN &   NaN &    NaN &           NaN &    NaN &    NaN &    NaN \\
			mba-lar-en-ro &           0.77 &   0.53 &   0.35 &  0.67 &    0.2 &  0.82 &     0.42 &   0.76 &   0.49 &         0.49 &      0.56 &  0.73 &  0.68 &     0.74 &   0.34 &  0.58 &    0.46 &   0.4 &   0.24 &          0.51 &   0.44 &   0.35 &   0.33 \\
			rob-bas &           0.77 &   0.48 &   0.28 &  0.51 &   0.26 &  0.59 &     0.36 &   0.76 &   0.53 &          0.5 &      0.59 &  0.75 &  0.74 &     0.71 &   0.41 &  0.53 &    0.51 &  0.46 &   0.16 &          0.55 &   0.42 &   0.47 &   0.46 \\
			rob-bas-ope-det &           0.77 &   0.28 &   0.32 &  0.45 &  0.095 &  0.54 &     0.23 &   0.71 &   0.35 &         0.35 &      0.56 &  0.75 &  0.63 &     0.72 &   0.14 &  0.54 &    0.32 &  0.36 &  0.082 &          0.51 &   0.31 &   0.38 &   0.35 \\
			rob-lar &           0.78 &   0.41 &   0.33 &  0.56 &  0.062 &  0.74 &     0.37 &   0.79 &    0.5 &         0.49 &      0.64 &  0.75 &  0.76 &     0.73 &   0.39 &  0.59 &    0.56 &  0.48 &   0.14 &           0.6 &   0.45 &   0.42 &   0.43 \\
			rob-lar-mnl &           0.79 &   0.32 &  0.073 &  0.47 &   0.31 &  0.83 &     0.02 &   0.76 &   0.47 &         0.49 &      0.61 &  0.75 &  0.72 &     0.73 &    NaN &  0.52 &    0.38 &  0.35 &   0.23 &          0.45 &   0.45 &   0.24 &   0.34 \\
			rob-lar-ope-det &           0.77 &   0.45 &   0.37 &   0.5 &   0.06 &  0.66 &     0.32 &   0.64 &   0.47 &         0.44 &      0.58 &  0.73 &  0.75 &     0.72 &   0.17 &  0.58 &    0.45 &   0.4 &   0.14 &          0.49 &    0.4 &   0.45 &   0.44 \\
			xlm-clm-end-102 &           0.69 &   0.31 &   0.35 &  0.35 &    0.2 &  0.47 &     0.21 &   0.68 &   0.32 &         0.51 &      0.55 &   0.6 &  0.67 &     0.74 &   0.27 &  0.57 &    0.38 &  0.44 &   0.18 &           0.5 &   0.34 &   0.28 &   0.38 \\
			xlm-clm-enf-102 &            0.7 &   0.22 &   0.17 &   0.3 &   0.25 &  0.48 &     0.29 &   0.61 &   0.32 &         0.29 &      0.45 &  0.56 &  0.66 &     0.72 &   0.24 &  0.49 &    0.39 &  0.32 &   0.11 &           0.4 &   0.38 &   0.39 &   0.31 \\
			xlm-mlm-100-128 &           0.77 &   0.27 &   0.33 &   0.2 &   0.26 &  0.54 &     0.35 &   0.69 &   0.49 &         0.48 &      0.57 &  0.66 &  0.72 &     0.68 &   0.25 &   0.5 &     0.4 &  0.39 &   0.22 &           0.5 &   0.34 &   0.44 &   0.15 \\
			xlm-mlm-17-128 &            0.7 &   0.26 &  0.036 &  0.32 &   0.29 &  0.56 &     0.39 &   0.69 &    0.5 &          0.4 &      0.53 &   0.7 &  0.69 &     0.75 &   0.32 &  0.55 &    0.46 &  0.39 &   0.19 &          0.47 &   0.39 &   0.38 &  0.085 \\
			xlm-mlm-en-204 &            0.8 &   0.42 &  0.067 &   0.6 &   0.33 &  0.86 &     0.17 &   0.71 &   0.47 &         0.54 &      0.57 &  0.73 &   0.7 &     0.67 &   0.39 &  0.56 &    0.54 &  0.39 &   0.23 &          0.54 &   0.48 &   0.43 &   0.41 \\
			xlm-mlm-end-102 &           0.74 &   0.35 &   0.26 &   0.4 &   0.24 &  0.59 &     0.32 &   0.72 &   0.46 &         0.49 &      0.56 &  0.62 &  0.64 &     0.71 &   0.22 &  0.49 &    0.43 &  0.37 &   0.22 &          0.52 &   0.29 &   0.41 &   0.38 \\
			xlm-mlm-enf-102 &            0.7 &   0.34 &   0.16 &  0.43 &    NaN &  0.58 &    0.095 &    0.6 &    0.4 &         0.45 &      0.53 &  0.62 &  0.67 &     0.68 &   0.34 &  0.56 &    0.38 &  0.41 &   0.16 &          0.53 &   0.33 &   0.35 &   0.39 \\
			xlm-mlm-enr-102 &           0.72 &   0.29 &   0.28 &  0.42 &   0.18 &  0.55 &     0.27 &    0.7 &   0.41 &         0.43 &      0.55 &  0.63 &  0.65 &     0.72 &   0.28 &  0.54 &    0.35 &   0.4 &   0.23 &          0.44 &   0.35 &   0.32 &   0.28 \\
			xlm-mlm-tlm-xnl-102 &            NaN &    NaN &    NaN &   NaN &    NaN &   NaN &      NaN &    NaN &    NaN &          NaN &      0.16 &   NaN &   NaN &      NaN &    NaN &   NaN &     NaN &   NaN &    NaN &           NaN &    NaN &    NaN &    NaN \\
			xlm-mlm-xnl-102 &           0.17 &    NaN &    NaN &  0.18 &   0.24 &   NaN &      NaN &    NaN &    NaN &          NaN &       NaN &  0.21 &  0.33 &      NaN &    NaN &   NaN &     NaN &  0.18 &  0.038 &           NaN &    NaN &    NaN &    NaN \\
			xlm-rob-bas &           0.75 &   0.34 &   0.34 &  0.21 &   0.09 &  0.53 &     0.15 &   0.75 &   0.31 &         0.52 &       0.6 &  0.71 &  0.67 &     0.69 &   0.16 &  0.53 &    0.46 &  0.45 &   0.19 &          0.58 &   0.35 &   0.39 &    0.4 \\
			xlm-rob-lar &           0.77 &   0.44 &   0.28 &  0.34 &   0.24 &  0.55 &     0.31 &   0.75 &   0.46 &         0.52 &      0.59 &  0.73 &  0.72 &     0.75 &   0.33 &   0.6 &     0.5 &  0.46 &   0.12 &          0.53 &   0.49 &   0.45 &   0.26 \\
			xlm-rob-lar-fin-con-eng &           0.77 &   0.33 &   0.38 &  0.33 &   0.17 &  0.73 &     0.26 &   0.74 &   0.39 &         0.34 &      0.57 &  0.71 &  0.74 &     0.72 &   0.32 &  0.57 &    0.46 &   0.4 &   0.15 &           0.5 &   0.48 &   0.38 &   0.38 \\
			xln-bas-cas &           0.75 &   0.37 &   0.38 &  0.37 &   0.12 &  0.54 &     0.31 &   0.81 &   0.52 &         0.49 &      0.59 &  0.68 &  0.74 &     0.73 &   0.38 &  0.58 &    0.54 &  0.46 &   0.19 &          0.47 &   0.41 &   0.43 &   0.37 \\
			xln-lar-cas &           0.78 &   0.38 &   0.36 &  0.39 &   0.26 &  0.65 &     0.43 &   0.79 &   0.55 &         0.53 &      0.56 &  0.73 &  0.75 &     0.75 &   0.34 &  0.58 &    0.55 &  0.52 &   0.28 &          0.41 &    0.4 &   0.43 &   0.32 \\
			
			\bottomrule
		\end{longtable}
	\end{landscape}	
	\restoregeometry

\end{document}